\documentclass[fleqn,usenatbib]{mnras}

\usepackage{newtxtext,newtxmath}

\usepackage[T1]{fontenc}

\DeclareRobustCommand{\VAN}[3]{#2}
\let\VANthebibliography\thebibliography
\def\thebibliography{\DeclareRobustCommand{\VAN}[3]{##3}\VANthebibliography}

\usepackage{graphicx}	
\usepackage{url}
\usepackage{amsmath}	
\usepackage{newtxtext,newtxmath}
\usepackage{xcolor}
\usepackage[normalem]{ulem}

\title[Statistics of onset temperature of solar flares]{Statistical analysis of the onset temperature of solar flares in 2010 - 2011}

\author[D. F. Silva et al.]{
Douglas F\'{e}lix da Silva,$^{1,2}$\thanks{E-mail: douglas93f@gmail.com}
Li Hui,$^{1,4}$
Paulo J. A. Sim\~{o}es,$^{2,3}$
Adriana Valio,$^{2}$ 
Costa, Joaquim E. R.,$^{5}$
\newauthor
Hugh S. Hudson, $^{3,6}$
Lyndsay Fletcher, $^{3,7}$
Laura A. Hayes $^{8}$ and
Iain G. Hannah $^{3}$
\\
$^{1}$ State Key Laboratory of Space Weather, National Space Science Center, Chinese Academy of Sciences, Beijing, China\\
$^{2}$Center for Radio Astronomy and Astrophysics Mackenzie, Engineering School, Mackenzie Presbyterian University, S\~ao Paulo, Brazil\\
$^{3}$SUPA School of Physics and Astronomy, University of Glasgow, Glasgow G128QQ, UK\\
$^{4}$University of Chinese Academy of Sciences, Beijing, China\\ 
$^{5}$ National Institute for Space Research (INPE), S\~ao Jose Dos Campos, Brazil\\
$^{6}$ Space Sciences Laboratory, UC Berkeley, 94720 CA, USA\\
$^{7}$ Rosseland Centre for Solar Physics, University of Oslo, PO Box 1029 Blindern, NO-0315 Oslo, Norway\\
$^{8}$ European Space Agency (ESA), European Space Research and Technology Centre (ESTEC), Keplerlaan 1, 2201 AZ Noordwijk, The Netherlands
}

\date{Accepted XXX. Received YYY; in original form ZZZ}

\pubyear{2022}

\begin{document}
\label{firstpage}
\pagerange{\pageref{firstpage}--\pageref{lastpage}}
\maketitle

\begin{abstract}

Solar flares are among the most energetic phenomena in the Solar System. Flare radiation, energetic particles, and associated coronal mass ejections are the main drivers of the space weather near the Earth. Thus, understanding the physical processes that trigger solar flares is paramount to help with forecasting space weather and mitigating the effects on our technological infrastructure. 
\citet{hudson_2021MNRAS.501.1273H} have recently identified a previously unknown phenomenon in solar flares: the plasma temperature, derived from soft X-ray (SXR) data, at the onset of four flares, was revealed to be in the range 10-15~MK, without evidence of gradual heating. To investigate how common the hot-onset phenomenon may be, we extend this investigation to  solar flares of B1.2~--~X6.9  classes recorded by the X-ray Sensor (XRS) on-board the GOES-14 and GOES-15 satellites between 2010 and 2011. 
For this statistical study, we employed the same methodology as \citet{hudson_2021MNRAS.501.1273H}, where the pre-flare SXR flux of each flare is obtained manually, and the temperature and emission measure values are obtained by the flux ratio of the two GOES/XRS channels using the standard software. 
From a total of 3224 events listed in the GOES flare catalog for 2010-2011, we have selected and analyzed 745 events for which the flare heliographic location was provided in the list, in order to investigate center-to-limb effects of the hot-onset phenomenon. 
Our results show that 559 out of 745 flares (75\%) exhibit an onset temperature above 8.6~MK (the first quartile), with respective  $\log_{10}$ of the emission measure values between 46.0 - 47.25 cm$^{-3}$, indicating that small amounts of plasma are quickly heated to high temperatures. These results suggest that the hot-onset phenomenon is very common in solar flares.
\end{abstract}




\section{Introduction} \label{sec:intro}

Solar flares and coronal mass ejections (CME) are the most energetic transient events observed in the solar atmosphere. Solar flares are rapid and intense brightness variations of solar emission detected in all electromagnetic spectrum, whereas CMEs consist of large amounts of magnetized plasma ejected into the interplanetary space. 

Flares occur within active regions in the solar atmosphere, and the released energy is believed to be of magnetic origin.  Due to magnetic reconnection, which facilitates the release of stored magnetic energy, flares accelerate particles, heat the local plasma, and produce radiation at all wavelengths \citep{benz2008,Fletcher2011SSRv..159...19F}.
Predicting when a major solar event will occur is crucial to mitigate the possible effects of space weather
on our planet. One way to do this is to 
better understand the causes of solar activity phenomena. Hence, studying the plasma conditions before the impulsive phase of a solar flare may clarify the physical processes that occur leading up to the main flare energy release.

\cite{hudson_2021MNRAS.501.1273H} analyzed a set of four flares observed in soft X-rays (SXR) by the X-ray Sensor (XRS) on-board the Geostationary Operational Environmental Satellite (GOES) and discovered that, at the onset of these events, the initial plasma temperature $T$ inferred from the SXR data was already high, around 10-15\, MK. The associated emission measure EM values were low (EM$< 10^{47}$ cm$^{-3}$), suggesting that a small volume of plasma was quickly heated to such high temperatures. The temperature and emission measure were inferred from the ratio of the flux between the two SXR channels, 1-8~\AA\ and 0.5-4~\AA, following \cite{GOES2005SoPh..227..231W}, and confirmed by X-ray spectroscopic analysis of Reuven Ramaty High Energy Solar Spectroscopic Imager \citep[RHESSI, ][]{RHESSI2002SoPh..210....3L} observations of the same events. Using ultraviolet images from the Atmospheric Imaging Assembly \citep[AIA, ][]{lemen_sdo} on board the Solar Dynamics Observatory \citep[SDO,][]{Pesnell2012}, the authors identified the location and size of the flare sources during this onset interval, and concluded that the 10-15\,MK plasma originated from very compact, low-lying, and short-lived features. 

Such hot, compact volumes of plasma have been previously observed both in SXR \citep{Hudson1994ApJ...422L..25H,Mrozek2004A&A...415..377M} and extreme ultraviolet (EUV) \citep{Fletcher2013ApJ...771..104F,Graham2013ApJ...767...83G,Simoes2015SoPh..290.3573S}, but during the impulsive phase of flares. During the impulsive phase, these compact sources are well-associated spatially with the flare ribbons, and display an impulsive behavior, before the coronal SXR and EUV emission\textbf{s} dominate. 

\textit{
 The immediately high temperature, between 10 and 15 MK, at the beginning of the event is characteristic of the hot-onset observed in the soft X-ray data. A gradual heating from the minimum detection of  GOES, of 4 MK, to the flare temperature to identify any period of. \cite{hudson_2021MNRAS.501.1273H} show that these compact events occur in the low solar corona. We do not know if this is common for all events. 
}

In this work we aim to verify how common the hot-onset phenomenon occurs in solar flares. For this, we analyze the events recorded by the GOES satellite between 2010 and 2011.

The data selection criteria and methodology are described in Section~\ref{sec:datamethod}. We discuss the results in Section~\ref{sec:results}. Finally, the main conclusions are presented in Section~\ref{sec:conclusion}.
\section{Temperature and Emission Measure at the onset of solar flares} \label{sec:datamethod}

\subsection{Data Selection}
\label{sec:data}
We analyzed solar flares that occurred between 2010 and 2011 and were detected in SXR by the GOES-14 and GOES-15 satellites. The standard routines to obtain and process GOES/XRS data, part of the Solar Software package \citep[SSW, ][]{SSW1998SoPh..182..497F}, a software library for solar analysis written in IDL, was used to obtain and reduce the data. To carry out this work, we used data from the Solar Data Analysis Center (SDAC) at NASA Goddard Space Flight Center.

In May 2022, a critical note
, regarding the temperature and emission measurements calculations from GOES XRS data, reports that the treatments of GOES-13 through 15 have been corrected due to an erroneous assumption about the 
calibration performed previously in SolarSoft. Thus any temperatures and emission measures calculated within SSW for GOES-13 through 17 before May 2022 are now considered incorrect by up to $30\%$. 
Our analysis was done before these changes, so we chose 
not to consider the changes to maintain exactly the same calibration as in 
\citep{hudson_2021MNRAS.501.1273H} for 
comparison. We warn that future work should take this into account.

Using the SSW routine \verb!get_gev.pro! we obtained the GOES flare list between 2010 and 2011 (the beginning of Cycle 24), with 3224 solar flares, including B-class events. 
The data, also obtained via SolarSoft tools, incorporate the recent GOES database recalibrations \citep{2021AGUFMSH25E2139P}.
The selection of the pre-flare time interval is usually performed by a researcher, a time-consuming method that hinders the statistical analysis of very large samples. In Sect.~\ref{sec:method} we discuss our approach to a straightforward and standardized background subtraction. We do note that an automated method was presented by \cite{TEBBS2012ApJS..202...11R}, the Temperature and Emission measure-Based Background Subtraction method (TEBBS). 
A comparison between our method and TEBBS is beyond the scope of the current paper and will be presented in the future.

In addition to characterizing the emission measure and temperature at the onset of the selected events, we also investigated whether center-to-limb effects are present. Therefore,
the heliographic location of each flare was also collected from the list. From the original 3224 events, 906 flares had a reported location. We further discarded 157 events, where the pre-flare flux of one of the channels (usually the high-energy one) was below the detection limit of the GOES/XRS.
Our final sample contained 745 events for the analysis: 6 X-class flares, 69 M-class, 524 C-class, and 146 B-class events.

\subsection{Background selection, Temperature and Emission Measure estimates}\label{sec:method}

To obtain the temperature $T$ and emission measure $EM$ at the onset of the selected events, only the excess flux emission during each solar flare's beginning must be considered. We manually carried out the pre-flare flux levels (background) subtraction procedure for each of the 745 analyzed events. This process consisted of two steps: (1) identify the start time of the solar flare and (2) select a time interval before the event's start. In the first step, we visually identified the beginning of each event; the parameters of the GOES list only helped us find the events in the observation light curve.
To obtain the background value, we use the average flux of the chosen background time interval, for each channel respectively.  

\begin{figure}
     \includegraphics[width=1.1\columnwidth]{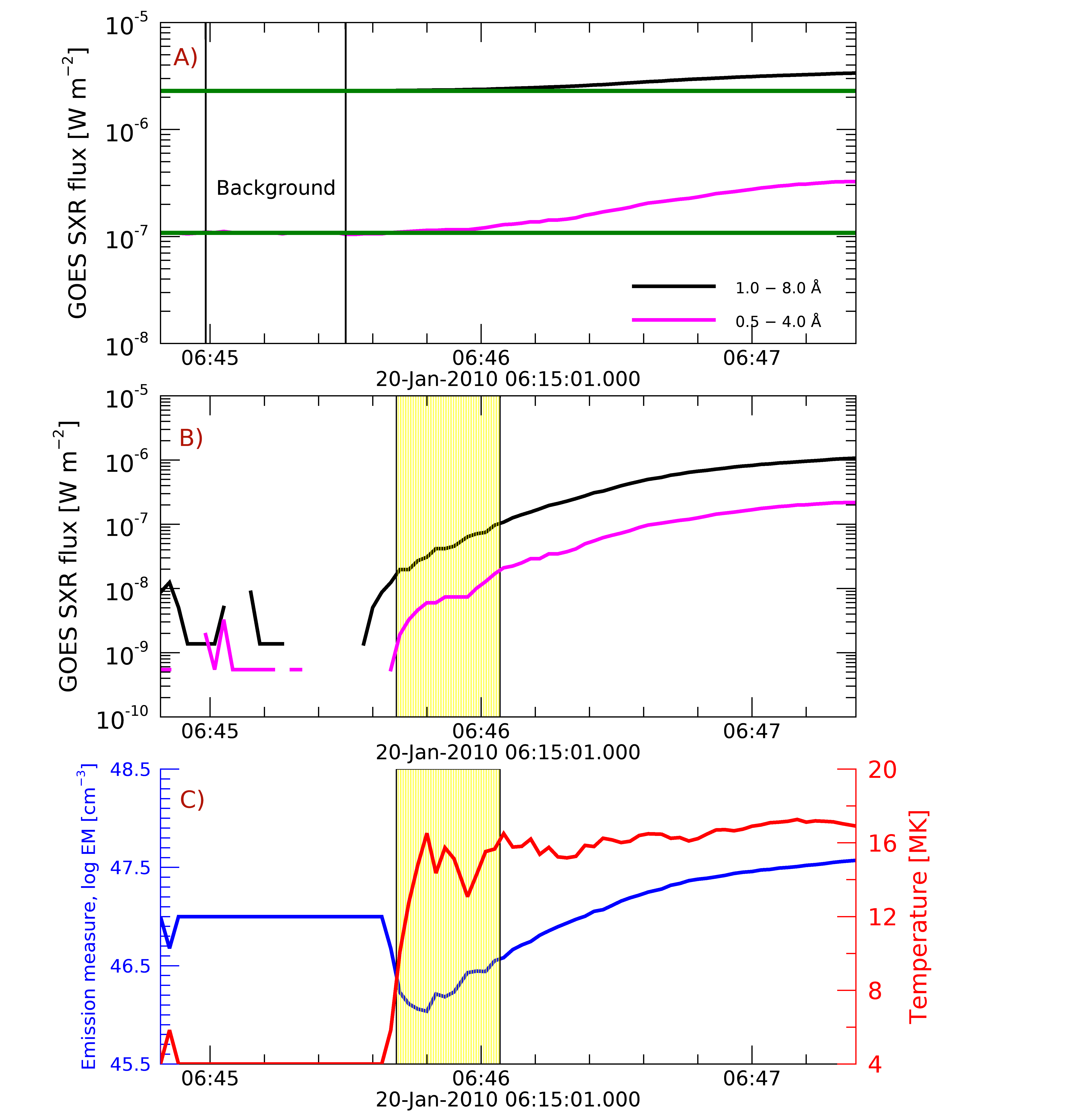}
     \caption{Initial phase of the SOL2010-02-20T06:45 event. \textbf{A)}  The GOES flux in the 1-8~\AA\ band (magenta curve) and in 0.5-4\AA\ (black curve) are shown. The vertical black lines limit the background time interval. The green horizontal lines depict the average flux estimated within the background time interval. \textbf{B)} The background subtracted flux in the 1-8~\AA\ band (magenta curve) and in 0.5-4\AA\ (black curve).
     The yellow strip is the interval chosen for the ``onset" interval, which in this case has a 20s duration.
    \textbf{C)}
     The calculated temperature and emission measure of the SXR source. The red curve represents the temperature (right y-axis) evolution during the flare, whereas the emission measure is shown in blue (left y-axis).
     }
     \label{fig:behavior}
\end{figure}

An example of the application of the methodology to obtain the $T$ and $EM$ of each flare is given in Figure~\ref{fig:behavior}. Figure~\ref{fig:behavior}a shows the initial phase of the SOL2010-02-20T06:45 event. The vertical lines mark the chosen pre-flare time interval with respect to the GOES flux for the channels 1-8~\AA~ (black line) and 0.5-4~\AA~ (magenta line); the green lines mark the average flux within the background window denoted in Figure~\ref{fig:behavior}a. The flux excess of both SXR channels are shown in Figure~\ref{fig:behavior}b, where we identify the flare onset: the instant where both SXR channels display values above the pre-flare flux (i.e. positive values). As discussed below, the onset interval, marked by the yellow region, was chosen to be 20 seconds.

Finally, Figure~\ref{fig:behavior}c shows the time evolution of the plasma temperature $T$ (red) and emission measure $EM$ (blue) as obtained via the standard routines \cite{GOES2005SoPh..227..231W}. The onset $T$ and $EM$ are calculated by taking the average of these quantities in the onset time interval. Note that the temperature is already above 10\,MK within the first twenty seconds of the event; this characterizes a hot-onset event. 

\begin{figure}
     \includegraphics[width=1.1\columnwidth]{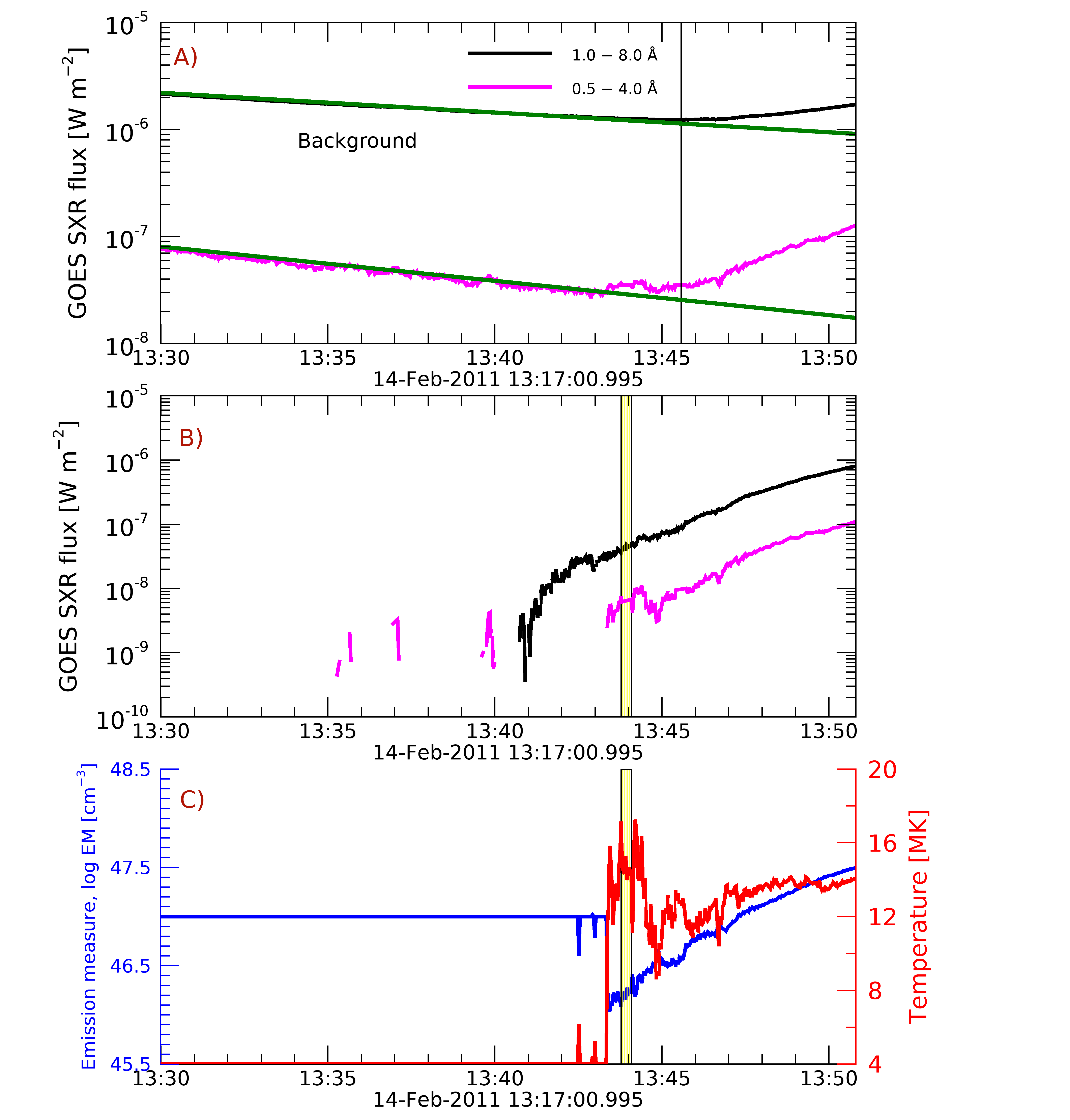}
     \caption{
     The initial phase of the SOL2011-02-14T13:44 event.
     The same as Figure~\ref{fig:behavior}, but with the background flux modeled as an exponential function. This was necessary because this flare occurred during the decay phase of a previous event.
     }
     \label{fig:behavior_exp}
\end{figure}

Some flares start during the gradual phase of a previous event. For these cases, the time variation of the pre-flare flux can be fitted by an exponential function (i.e. a straight line in log space), under the assumption that this model captures both the flux of the gradual phase of the previous event and the flux contribution from the rest of the Sun. Note that we only applied the exponential fit method to flares that occurred during the gradual phase of a previous event.
Figure~\ref{fig:behavior_exp}a shows an example of such a case, SOL2011-02-14T13:17, where the flux in the background interval for each GOES/XRS channel is fitted with an exponential function (green line). Figure~\ref{fig:behavior_exp}b and \ref{fig:behavior_exp}c show the excess flux and the resulting $T$ and $EM$, the same as Figure~\ref{fig:behavior}. It was necessary to employ this procedure for 130 events (of out 745) in our sample.

\begin{figure}
     \includegraphics[width=1.1\columnwidth]{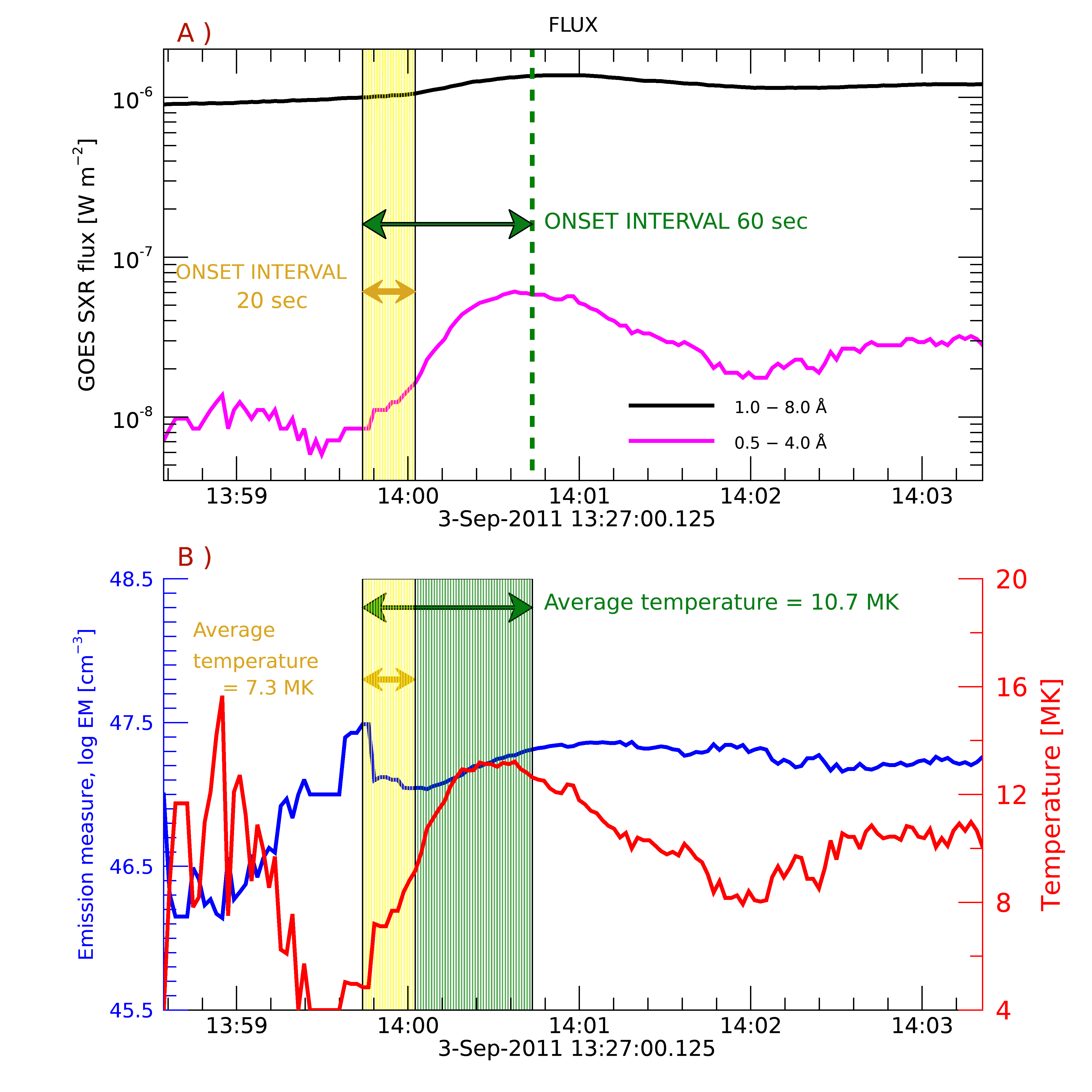}
     \caption{Comparison of onset time intervals of 20 and 60 seconds (indicated by the horizontal arrows), for the event SOL2011-09-03T13:27UT. (a) SXR flux for the 1-8~\AA\ band (black curve) and in 0.5-4\AA\ (magenta curve) both with background and (b) temperature $T$ and emission measure $EM$ derived from the SXR data. Note that the 60~s interval captures the instant of a maximum flux of the event  and the peak temperature, indicating that this choice of onset interval is inadequate. The average temperature in the 60-second interval (10.7 MK) is higher than that of the 20-second interval (7.3 MK) because, for short events, the 60-second interval includes the start event to the peak.}
     \label{fig:hot_20_60_sec}
\end{figure}

\begin{figure}
     \includegraphics[width=1.0\columnwidth]{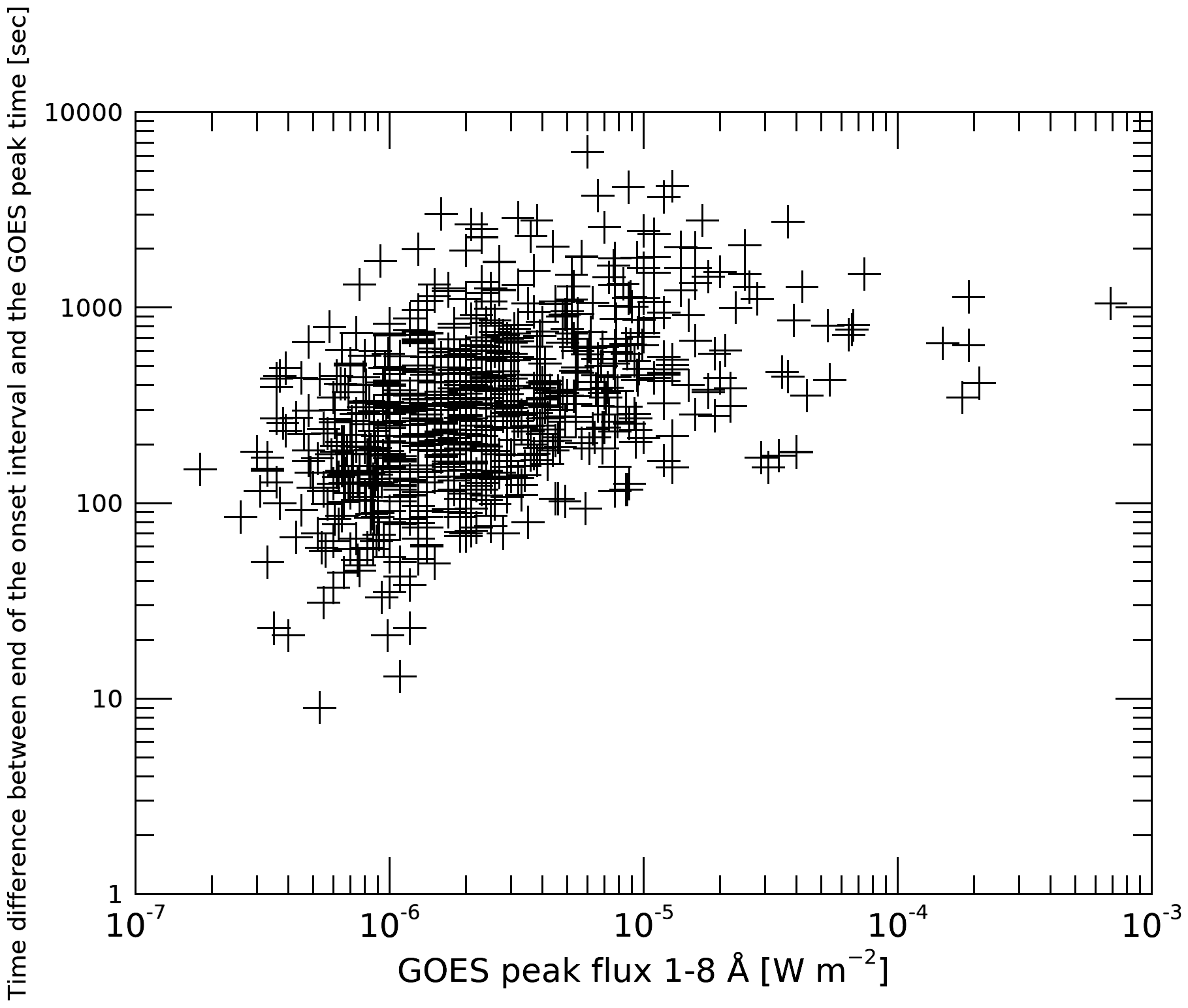}
     \caption{Time difference between the end of the selected onset time of 20 seconds and the GOES peak time for all analyzed flares. }
     \label{fig:deltatime}
\end{figure}

We decided to adopt a fixed onset interval of 20 seconds. The choice for the onset time interval is somewhat arbitrary but a necessary one for a robust statistical analysis. With a 20 s interval, there is enough data for averaging the $T$ and $EM$ values, with the XRS cadence of 2 seconds. With longer onset intervals, the rising phase and/or even the event's maximum might be captured. As an example, we compare two onset intervals: 20 and 60 seconds, shown in Figure~\ref{fig:hot_20_60_sec}. The two intervals refer to a time interval measured forward from the start of the flare, as defined previously. Note that the 60-second interval (indicated by the green arrow) spans the beginning of the event until past its maximum. The resulting onset temperatures are $T(20s)=7.3$~MK and $T(60s)=10.7$~MK, with the latter including the maximum temperature of the event. Thus, for short-duration events, a 60-second onset interval is too long. Therefore, a 20-second time interval is more appropriate for a statistical approach. 
We also checked whether the selected onset interval for each flare was sufficiently separated from the SXR peak time. In Figure \ref{fig:deltatime} we show the time difference between the end of the selected onset interval to the SXR peak time, for all analyzed events. This time separation is well above tens of seconds for most events. All results showed hereafter were produced adopting an interval of 20 seconds for the onset.

\section{Results}\label{sec:results}

\subsection{Statistics of the onset T and EM}

\begin{figure}

     \includegraphics[width=\columnwidth]{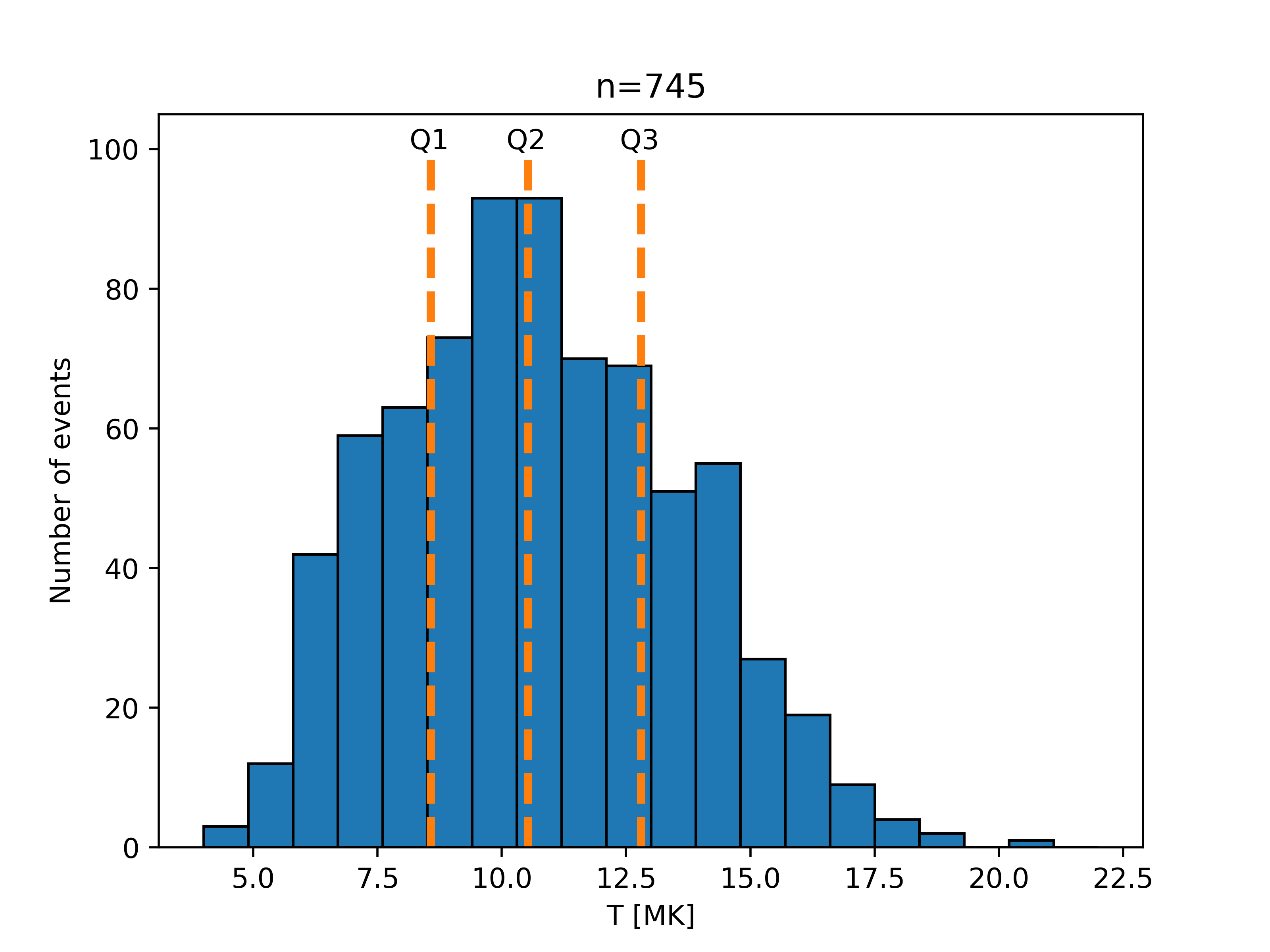}
     \includegraphics[width=\columnwidth]{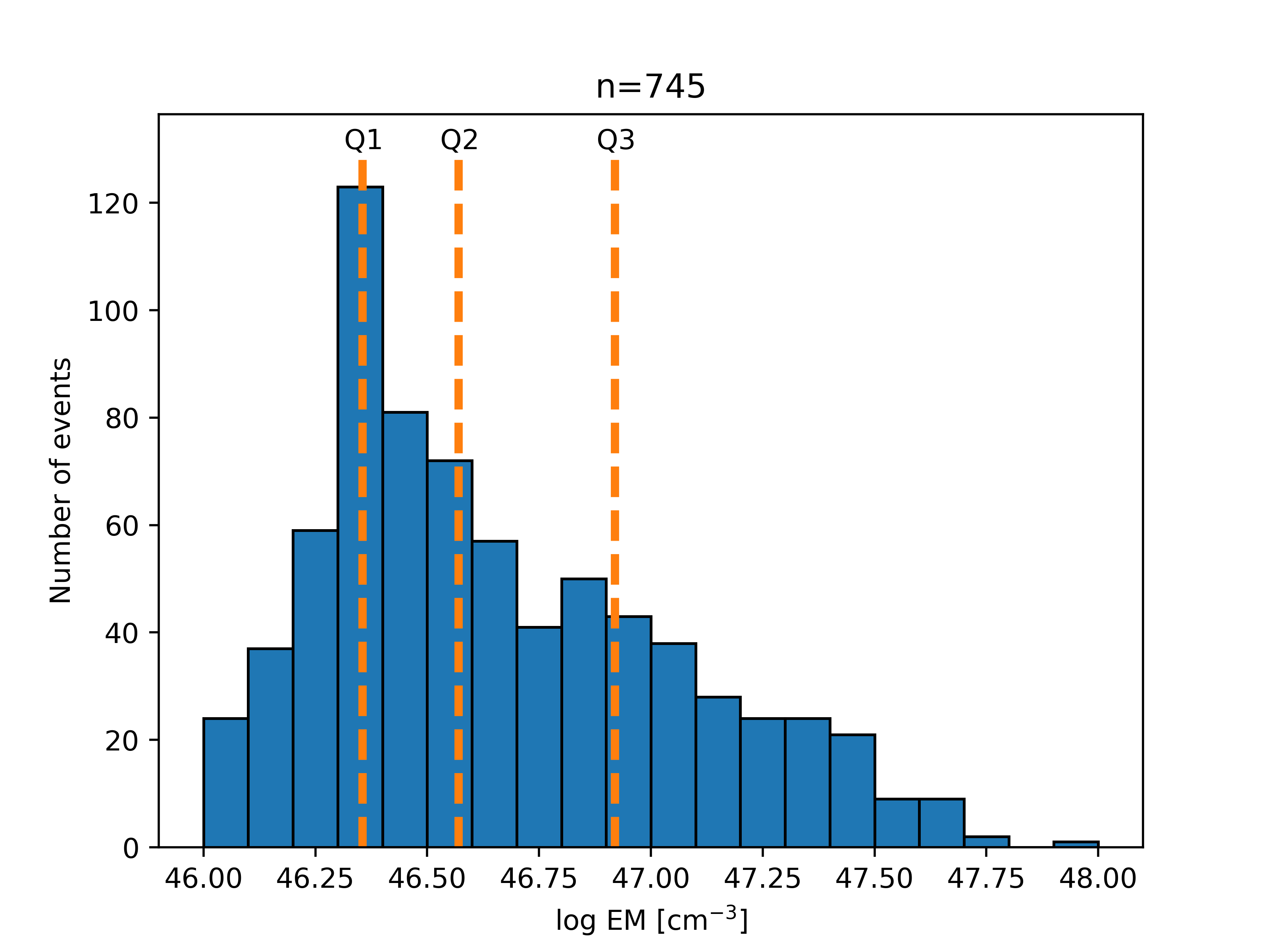}
      \caption{Distributions of the onset temperature (top panel) and emission measure (bottom panel) for the 745 flares in the period 2010-2011, considering a 20~s onset time interval. The vertical lines show the quartile values.}  
     \label{fig:histo_temp_em}
\end{figure}

The temperature and the emission measure for all 745 events were calculated considering an onset interval of 20~s. Histograms of the onset temperature and the emission measure for all 745 events are shown in Figure~\ref{fig:histo_temp_em} in the top and bottom panels, respectively. The $T$ distribution has a clear maximum of around 10~MK, and it is roughly symmetrical. The $EM$ distribution also has a clear maximum, near $10^{46.3}$~cm$^{-3}$, but shows a longer tail towards higher values. We computed the quartiles of each distribution to quantify the interpretation of our results; these are indicated by the vertical lines in each panel of Figure~\ref{fig:histo_temp_em} and are presented in the Table~\ref{tab:histo_temp}.

We find that only $\approx 25$\% of our flare sample (186 events) have an onset temperature below 8.6~MK. The majority of the distribution presented onset temperatures that characterize a hot onset: 373 events with $8.6\mathrm{MK} < T < 12.8\mathrm{MK}$ and 186 events with $T > 12.8$\,MK. These results show that 75\% of our sample (559 of 745 events) have an onset temperature above 8.6~MK. The sample presents a median temperature of 10.5~MK, within the initial 20 seconds from detecting SXR flux excess.

For the $EM$ quartile analysis, 187 of the events have EM < $10^{46.4}$cm$^{-3}$, 412 events with 46.4 < log EM < 47.0, and 146 flares with log EM > 47.0. These results indicate that only small amounts ($EM < 10^{48}$ cm$^{-3}$) of plasma are heated during the flare onset, as expected.

\begin{table}
\centering
\caption{Quartiles of the onset temperature and emission measure for 745 events.}
\begin{tabular}{cccc}
Quartiles & Q1 & Q2 & Q3 \\
\hline
$T$ (MK) & 8.6 & 10.5 & 12.8 \\ \hline
Number of events  & 186     & 373        & 559       \\ 
\hline \hline
$\log EM$ (cm$^{-3}$) & 46.4 & 46.6 & 47.0 \\ 
Number of events  & 187  & 374   & 599 
\end{tabular}
\label{tab:histo_temp}
\end{table}

\begin{figure}
\centering

    \includegraphics[width=\columnwidth]{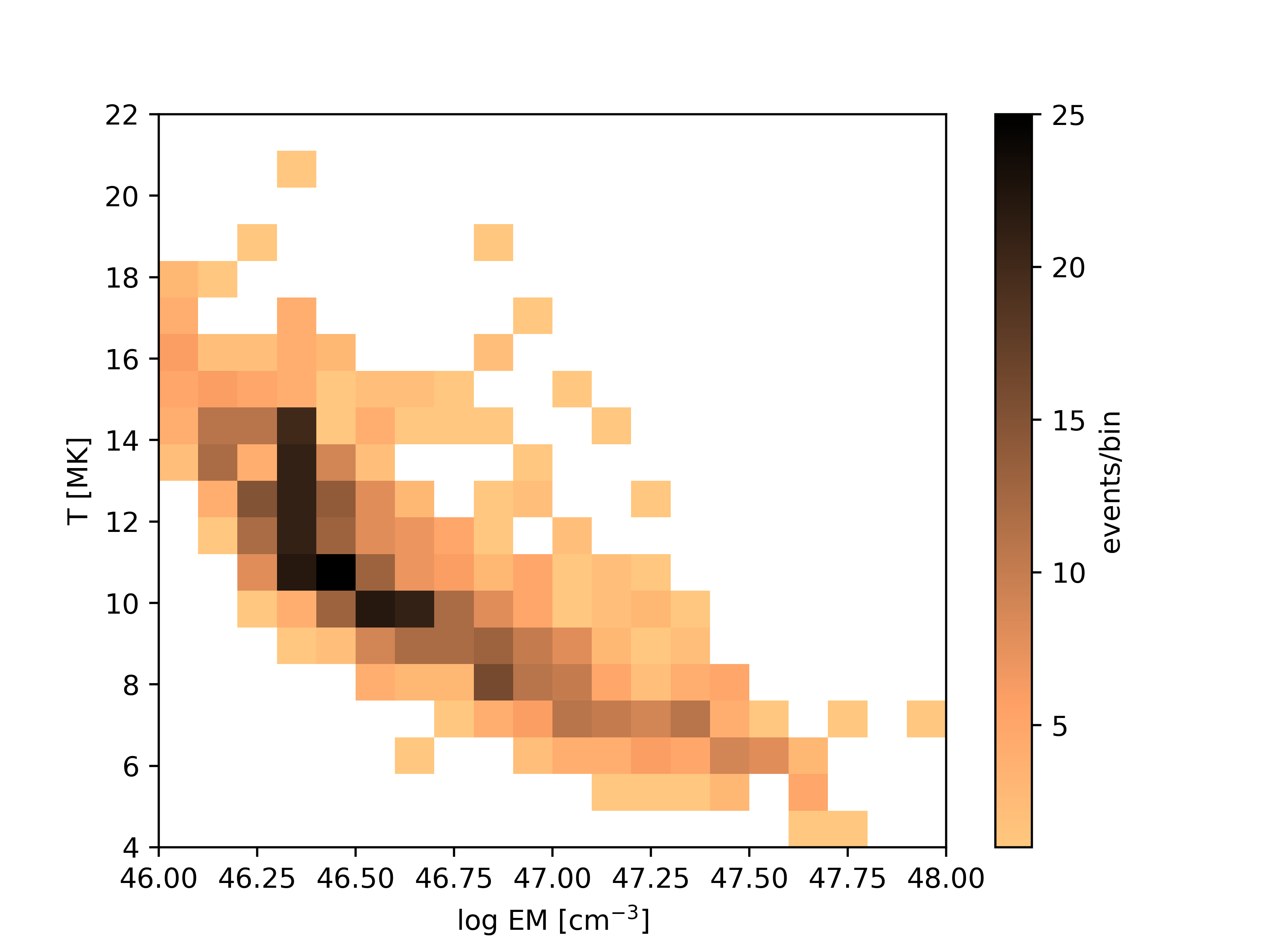}
     \caption{2D histogram of the average temperature and emission measure estimated during the 20s onset interval for the 745 analyzed events. 
     }
     \label{fig:tem-em}
\end{figure}

The correlation between temperature and emission measure at the onset of all the flares in our sample is shown in Figure~\ref{fig:tem-em}, which shows a 2D histogram of the distribution of $EM$ and $T$. We see an anti-correlation between these quantities for the events, where the higher the emission measure, the lower the temperature. Most of the events are clustered between temperatures of $10-15$ MK with EM $\sim 1.8-5.6 \times 10^{46}$ cm$^{-3}$.

\subsection{Center-to-limb effects on the onset T and EM}

\cite{hudson_2021MNRAS.501.1273H} show, in their Figure 6, the onset temperature of flares occurring on the Active Region NOAA 11748 during its appearance on the eastern limb. They conclude that the solar limb partially occults the flares before mid-day May 2013. Thus, the onset temperature (around or above 20\,MK) is not the \textit{true} temperature of the onset source (originating from compact, low-lying sources), but reflects the temperature of the emission from the coronal loops well into the evolution of the flare. As the AR rotates fully into view of the GOES/XRS sensors, the \textit{true} onset temperatures are visible, yielding temperatures in the range of 10--15\,MK.

\begin{figure}
\centering
        \includegraphics[width=\columnwidth]{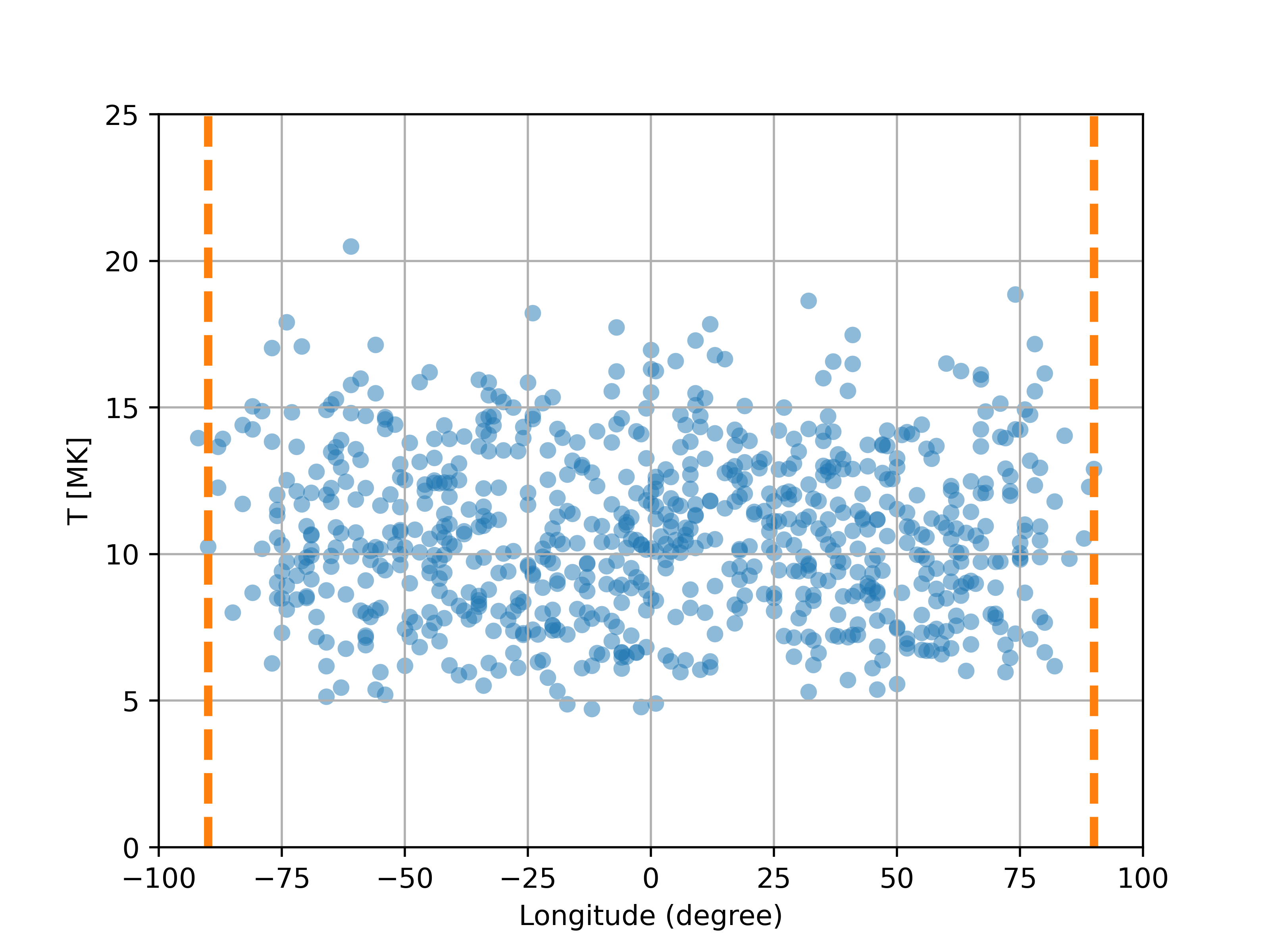}
     \caption{Average temperature during the onset of solar flares as a function of their longitude, as listed in the GOES catalog. 
    The orange vertical dashed lines mark the -90$^\circ$ and 90$^\circ$ maximum longitudes.  
     }
     \label{fig:longitude}
\end{figure}

To investigate this phenomenon, and any other center-to-limb effects, we also analyze the onset temperature as a function of the solar longitude of each flare, as shown in Figure~\ref{fig:longitude}. No correlation is seen between temperature and location of the flare, that is, no center-to-limb variation is observed.

The four events analyzed by Hudson et al. 2021 showed onset temperatures between 10 and 15MK. The $T>20$~MK onsets refer to the events of a single active region, as it rotated into view, in May 2013. We do not find similar temperatures in our sample of events; however, we also do not have a case of a limb-occulted flare, judging by the flare location data. As suggested by Hudson et al., the higher onset temperatures can be explained by detecting the SXR emission from the higher flares loops, already in an advanced phase of the flares. At the same time, the solar limb obscured the actual onset.

\section{Concluding remarks}\label{sec:conclusion}

\cite{hudson_2021MNRAS.501.1273H} revealed the existence of high temperature (10--15~MK) plasma, in small amounts ($EM < 10^{47}$ cm$^{-3}$), at the onset of four flares, ranging from classes B to M. These properties offer a new challenge for standard models and theoretical studies of the physics of solar flares. 

To verify how common this newly revealed phenomenon is, we performed a statistical analysis of the temperature $T$ and emission measure $EM$, obtained from the standard methodology from GOES X-ray Sensor data, for flares in the period 2010-2011. GOES-14 and GOES-15 satellites observed over three thousand flares during this period. From this sample, we have analyzed 745 events that had recorded, in addition to the GOES class, the start, peak, and end times, also the location and identification of the active region where the event occurred. This spatial information was important to locate the onset source on the solar disk.

Our results from a quartile analysis show that only 25\% of the flare sample had an onset temperature below 8.6~K, therefore, 75\% of our sample can be characterized as having a hot onset. The distribution of the onset temperatures for the 745 flares is unimodal, with a median of 10.5~MK. Also, there is a continuous distribution between the emission measure and temperature (see Figure~\ref{fig:tem-em}), where hotter events show a small contribution from the emission measure, and events with a lower temperature exhibit a larger emission measure value.
The shape of this distribution results at least partially from the nonlinear correlation between parameter uncertainties, which are roughly connected by $I \propto EM \times e^{-\mathrm{const.}/T}$.
The emission measures are generally tiny relative to their peak values, and the 20-s snapshots at the earliest times in the events have limited signal-to-noise ratios.
The values in Figure~\ref{fig:tem-em} suggest that, for most cases, small amounts of plasma are heated almost immediately to temperatures around 10~MK. Future studies with higher temporal and spatial resolutions may bring further information regarding the properties of the hot onset of flares. Only for 25\% of the flares in our sample is the initial temperature below 8.6~MK, where gradual heating can be observed during the initial phase of the events. Also, we find no center-to-limb dependence of the onset temperature (Figure ~\ref{fig:longitude}).

Our study strongly indicates that the hot-onset phenomenon, revealed by \cite{hudson_2021MNRAS.501.1273H}, is a common feature of solar flares. Given the possible thermal nature of the hot onsets \citep{hudson_2021MNRAS.501.1273H}, we anticipate that new clues might be obtained via observations in the infrared/THz range with the 30~THz cameras AR30T \citep{Lopez2022A&A...657A..51L}, BR30T \citep{2018SpWea..16.1261G} and the new 15~THz \textit{High Altitude THz Solar Telescope} \citep[HATS, ][]{HATS2020SoPh..295...56G}, soon to be operating at the Observatorio Astronómico Félix Aguilar (OAFA), at the Argentinean Andes.

\section*{Acknowledgements}

DFS acknowledges the support from the China-Brazil Joint Laboratory for Space Weather (CBJLSW). HL is supported by NNSFC grants (42022032, 41874203, 42188101), International Partnership Program of CAS (Grant No. 183311KYSB20200017). PJAS acknowledges support from the Fundo de Pesquisa Mackenzie (MackPesquisa) and CNPq (contract 307612/2019-8). This research was partially supported by FAPESP grant 2013/24155-3. LF acknowledges support from the UK's Science and Technology Facilities Council (grant number ST/T000422/1). LF and IGH acknowledge support from the UK's Science and Technology Facilities Council (grant number ST/T000422/1). L.A.H is supported by an ESA Research Fellowship.

\section*{DATA AVAILABILITY}
All data used in this work reside in the public domain at \url{https://umbra.nascom.nasa.gov/goes/fits/}



\bibliographystyle{mnras}
\bibliography{CBJLSW} 








\bsp	
\label{lastpage}
\end{document}